\documentclass[3p,twocolumn,authoryear]{elsarticle}
\usepackage{graphicx}
\usepackage{epsf}
\usepackage{wrapfig}
\usepackage{epsfig}
\usepackage{xspace}
\usepackage{ulem}

\def\b0{{\mbox{\boldmath$0$}}}

\usepackage{morefloats}
\usepackage{amsfonts}
\usepackage{color}
\usepackage{graphicx}
\usepackage{epsf}
\usepackage{lineno}
\usepackage{wrapfig}
\usepackage{epsfig}
\usepackage{sublabel}
\usepackage{epsfig}

\def\beq{\begin{equation}}
\def\eeq{\end{equation}}

\def\beqy{\begin{eqnarray}}
\def\eeqy{\end{eqnarray}}

\mathchardef\mhyphen="2D
\newcommand{\eg}{e.g.\xspace}
\newcommand{\ie}{i.e.\xspace}

\def \b #1{ {\bf #1}}
\newcommand{\be}{\begin{eqnarray}}
\newcommand{\ee}{\end{eqnarray}}

\def \b #1{ {\bf #1}}

\def \b #1{ {\bf #1}}
     \font\tenbifull=cmmib10 scaled 1200 % bold math italic
     \font\tenbimed=cmmib9
     \font\tenbismall=cmmib7
\mathchardef\bbkappa="7114
\mathchardef\bbrho="711A
\mathchardef\bbsigma="711B
\mathchardef\bbtau="711C
\mathchardef\bbvarrho="7125
\mathchardef\bbvarsigma="7126
\mathchardef\bbxi="7118

\makeatletter
\long\def\@makecaption#1#2{%
  \vskip\abovecaptionskip
  \sbox\@tempboxa{#1. #2}%
  \ifdim \wd\@tempboxa >\hsize
    #1. #2\par
  \else
    \global \@minipagefalse
    \hb@xt@\hsize{\hfil\box\@tempboxa\hfil}%
  \fi
  \vskip\belowcaptionskip}
\makeatother
\begin{document}
\begin{frontmatter}
\vskip 2mm \date{\today}\vskip 2mm
\title{Scaling properties of rainfall induced landslides predicted by a physically based model}
\author[cnr]{Massimiliano Alvioli}
\author[cnr]{Fausto Guzzetti$^\star$}
\author[cnr,unipg]{Mauro Rossi}
\address[cnr]{Consiglio Nazionale delle Ricerche, Istituto di Ricerca per la Protezione Idrogeologica,
  via Madonna Alta 126, I-06128 Perugia, Italy}
\address[unipg]{Universit\`a degli Studi di Perugia, Dipartimento di Scienze della Terra,\\ 
  Piazza Universit\`a, I-06123, Perugia, Italy}
\vskip 2mm
%------------------------------------------------------------------------------%
%------------------------------------------------------------------------------%
%\maketitle
\begin{abstract}\label{sec:abstract}
Natural landslides exhibit scaling properties revealed by power law relationships.
These relationships include the frequency of the size (\eg, area, volume) of the 
landslides, and the rainfall conditions responsible for slope failures in a region.
Reasons for the scaling behavior of landslides are poorly known.
We investigate the possibility of using the Transient Rainfall Infiltration and Grid-Based Regional 
Slope-Stability analysis code (TRIGRS), a consolidated, physically-based, numerical 
model that describes the stability/instability conditions of natural slopes forced 
by rainfall, to determine the frequency statistics of the area of the unstable slopes 
and the rainfall intensity $(I)$ -- duration $(D)$ conditions that result in landslides 
in a region. We apply TRIGRS in a portion of the Upper Tiber River Basin, Central Italy. 
The spatially distributed model predicts the stability/instability conditions of individual 
grid cells, given the local terrain and rainfall conditions. We run TRIGRS using multiple, 
synthetic rainfall histories, and we compare the modeling results with empirical evidences 
of the area of landslides and of the rainfall conditions that have caused landslides in the 
study area. Our findings revealed that TRIGRS is capable of reproducing the frequency of 
the size of the patches of terrain predicted as unstable by the model, which match the 
frequency size statistics of landslides in the study area, and the mean rainfall $D,I$ 
conditions that result in unstable slopes in the study area, which match rainfall 
$I\mhyphen D$ thresholds for possible landslide occurrence. Our results are a step 
towards understanding the mechanisms that give rise to landslide scaling properties.
\end{abstract}
\end{frontmatter}
% == INTRODUCTION ============================================================ 
\section{Introduction}\label{sec:intro}

There is accumulating evidence that natural landslides exhibit scaling properties 
\citep{Hergarten:2000,Hergarten:2002,Turcotte:2002,Chen:2011}, including the area and  
volume of the slope failures 
\citep{Pelletier:1997,Stark:2001,Guzzetti:2002, Malamud:2004,VanDenEeckhaut:2007,Brunetti:2009a}, 
and the amount of rainfall required for the initiation of landslides in a region
\citep{Caine:1980,Innes:1983,Aleotti:2004,Guzzetti:2007,Guzzetti:2008a}.
The scaling properties of landslides are revealed by power law dependencies, and are 
considered evidence of the critical state of landscape systems dominated by slope wasting 
phenomena \citep{Hergarten:2002,Turcotte:2002}

It is known that, regardless of the physiographic or the climatic settings, the probability 
(or frequency) density of  event landslides increases with the area of the landslide up to 
a maximum value, known as the ``rollover", after which the density decays along a power law
\citep{Stark:2001,Malamud:2004,VanDenEeckhaut:2007}. 
The length scale for the rollover, and the rapid decay along a power law, are conditioned by 
the mechanical and structural properties of the soil and bedrock where the landslides occur 
\citep{Katz:2006,Stark:2009}, and are independent of the landslide trigger \citep{Malamud:2004}.
The probability (or frequency) density of the landslide volume obeys a negative power-law, 
with a scaling controlled by the type of the landslides \citep{Brunetti:2009a}.
The dependence of landslide volume on landslide area was also shown to obey a distinct scaling 
behavior over more than eight orders of magnitude \citep{Guzzetti:2009,Larsen:2010,Klar:2011}.

Rainfall is a recognized trigger of landslides, and early investigators have recognized that
empirical rainfall thresholds can be established to determine the amount of rainfall required 
to initiate landslides in a region 
\citep{Endo:1970,Caine:1980,Govi:1980,Innes:1983,Moser:1983,Cancelli:1985}. 
Different types of empirical thresholds that use combinations of rainfall measurements obtained 
from the analysis of rainfall events that resulted (or did not result) in landslides were proposed 
in the literature, including mean intensity - duration ($I\mhyphen D$) and rainfall cumulated 
event - duration ($E\mhyphen D$) thresholds, and their variations \citep{Guzzetti:2007,Guzzetti:2008a}. 
With a few exceptions \citep{Wieczorek:1987,Cannon:1985,Crosta:2003} all the empirical rainfall 
thresholds are represented by power law models, indicative of the self-similar behavior of the 
rainfall characteristics responsible for landslide occurrence.

Despite the abundant empirical evidence, the reasons for the scaling behaviors of landslide 
phenomena are poorly known, and only a few attempts were made to interpret the empirical evidences 
with deterministic or physically based models \citep{Katz:2006,Stark:2009}.
In this paper, we show that a relatively simple, physically based model that describes the 
stability/instability conditions of slopes forced by rainfall, when applied to a sufficiently 
large geographical area produces results that are in agreement with two known scaling properties 
of landslides, namely: (i) the rainfall conditions that result in unstable slopes, which match
regional empirical $I\mhyphen D$ thresholds for possible landslide occurrence 
\citep{Guzzetti:2007,Guzzetti:2008a}, and (ii) the frequency distribution of the area of the 
patches of terrain predicted as unstable by the model, which matches the statistics of landslide 
area for event landslides \citep{Pelletier:1997,Stark:2001,Malamud:2004,VanDenEeckhaut:2007}.

The paper is organized as follows. In Section \ref{sec:data}, we describe the geographical 
area in Central Italy where we have conducted our experiments (Fig. \ref{fig01}), and in 
Section \ref{sec:model} we provide general information on the Transient Rainfall Infiltration 
and Grid-Based Regional Slope-Stability analysis code (TRIGRS, version 2.0; \citealt{Baum:2008}) 
that we adopted for the experiments. Next, in Section \ref{sec:iddep}, we compare the $I\mhyphen D$
conditions capable of producing slope instability in the study area predicted by TRIGRS, with 
empirical rainfall $I\mhyphen D$ thresholds for possible landslide occurrence in Central Italy.
This is followed by a comparison of the probability density of the area of the patches of terrain 
predicted as unstable by TRIGRS in the study area with the probability density of natural landslides 
in the same general area, and by a discussion about the possible relations of this work to other existing 
approaches for the description of landslide scaling phenomena. We conclude, in Section \ref{sec:concl}, 
summarizing the lessons learnt.

% == STUDY AREA ============================================================ 

\section{Study area and data}\label{sec:data}

\begin{figure}[!htp]
  \vskip -0.5cm
  \centerline{\hspace{0cm}
    \epsfxsize=0.35\textwidth\epsfbox{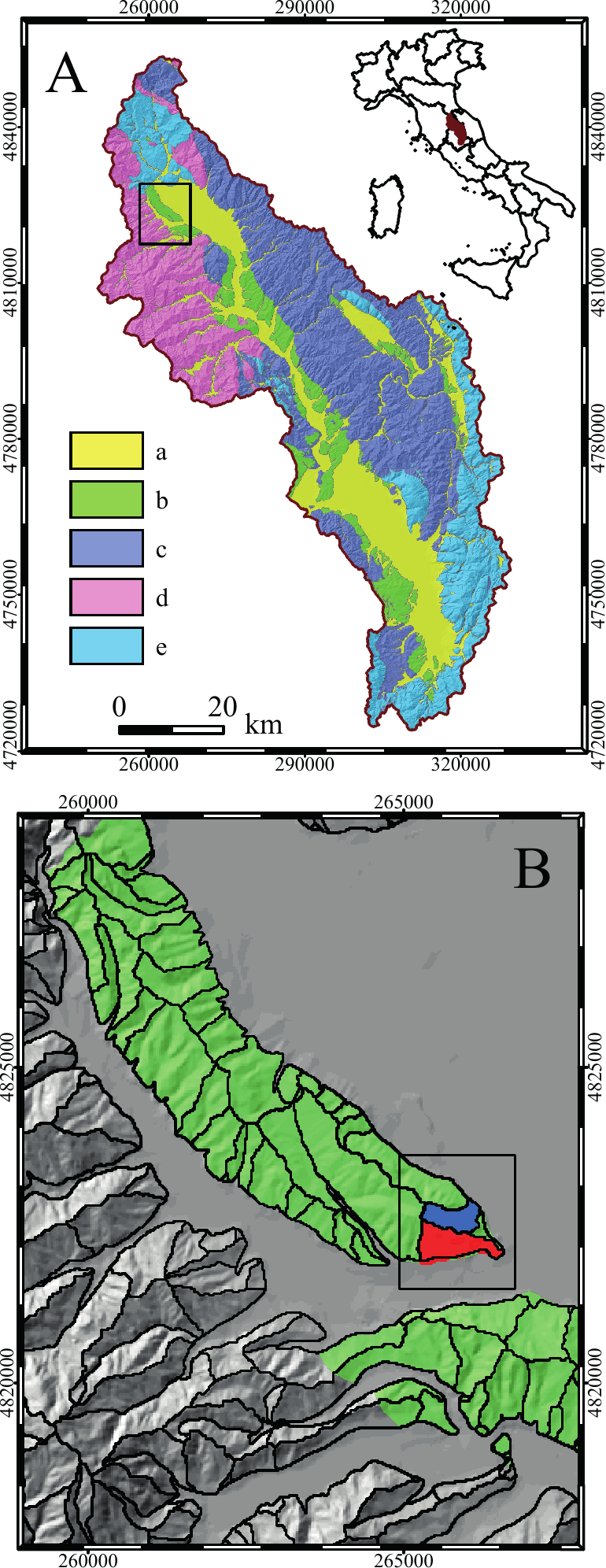}}
 \caption{
  The Upper Tiber River Basin (UTRB), in Central Italy. 
    Colours in (A) show five lithological complexes
    \citep{Cardinali:2001,Guzzetti:2008b}:
   (a) Fluvial and lake deposits, recent in age. 
   (b) Continental, post-orogenic sediments, Pliocene to Pleistocene.
   (c) Allochthonous rocks, lower to middle Miocene.
   (d) Tuscany and Umbria turbidites sequences, Eocene to Miocene.
   (e) Umbria-Marche sedimentary sequence, Lias to lower Miocene.
    Black box in (A) shows location of enlargement portrayed in (B) where
    black lines show hydrological sub-basins derived automatically from a 
    25 $\times$ 25 m DEM; the only lithological area shown in colour in (B) is b of (A),
    and in (B) two sample sub-basins are shown in red and blue.
    Black box in (B) shows location of Fig. \ref{fig02}.
  }
  \label{fig01}
\end{figure}

The Upper Tiber River basin (UTRB) extends for 4,098 km$^2$ in Central Italy, with elevation 
in the range from 163 m at the basin outlet to 1571 m along the divide between the Adriatic 
Sea and the Tyrrhenian Sea (Fig. \ref{fig01}). In the area the landscape is hilly or mountainous, 
with open valleys and intra-mountain basins. In the mountains and the hills, the morphology is 
conditioned by lithology and the attitude of the bedding planes. Climate is Mediterranean, with 
most of the precipitation falling from October to December and from February to April 
\citep{Cardinali:2001,Guzzetti:2008b}. 

For the UTRB two digital representations of the terrain elevation (DEM) were available to us. 
A coarser DEM, with a ground resolution of 25 $\times$ 25 m, was obtained through the linear 
interpolation of elevation data along contour lines shown on 1:25,000 topographic base maps 
\citep{Cardinali:2001}. A finer DEM, with a ground resolution of 10 $\times$ 10 m, was prepared 
by the Italian National Institute for Geophysics and Volcanology through the interpolation of 
multiple sources of elevation data \citep{Tarquini:2007,Tarquini:2012}. 

Five lithological complexes, or groups of rock units, crop out in the UTRB \citep{Cardinali:2001,Guzzetti:2008b}, 
including, from younger to older: 
(a) recent fluvial and lake deposits, which crop out mostly along the valley bottoms, 
(b) unconsolidated and poorly consolidated sediments pertaining to a continental, 
post-orogenic sequence, Pliocene to Pleistocene in age, 
(c) allochthonous rocks, lower to middle Miocene in age, 
(d) sediments pertaining to the Tuscany turbidites sequence, Eocene to Miocene in age, 
and to the Umbria turbidites sequence, Miocene in age, and
(e) sediments pertaining to the the Umbria-Marche stratigraphic sequence, Lias to lower Miocene in age. 
% ======================================================================== Fig ONE
%% =================================================================== End Fig ONE
Soils reflect the lithological types, and range in thickness from less than 20 cm to more than 1.5 m. 
Landslides are abundant in the area, 
% ======================================================================== Fig TWO
\begin{figure}[!htp]
  \vskip -0.5cm
  \centerline{\hspace{0cm}
    \epsfxsize=0.35\textwidth\epsfbox{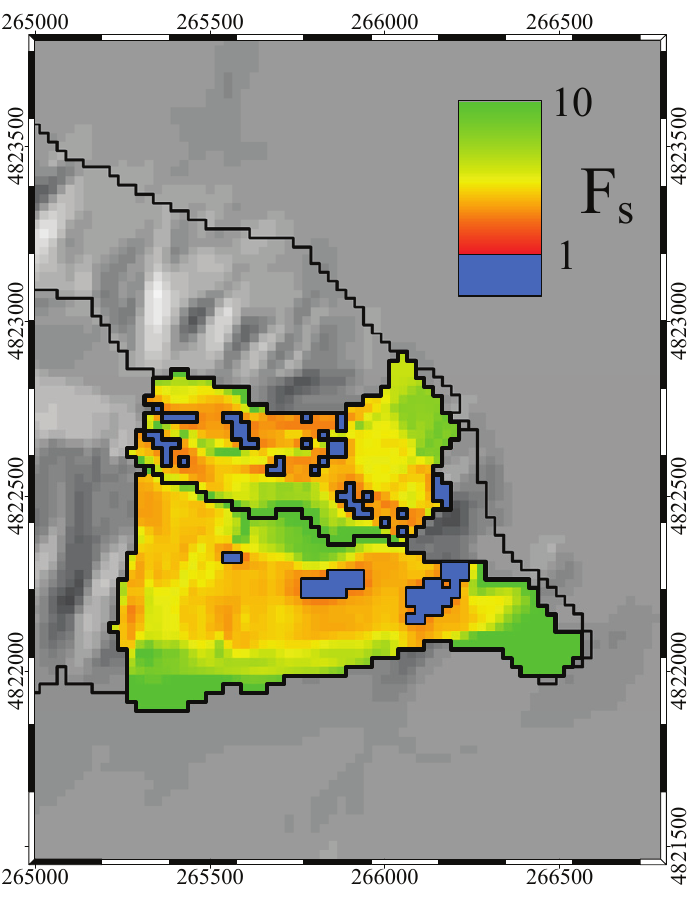}}
  \caption{
    Result of the TRIGRS numerical modeling for two sub-basins where
    unconsolidated and poorly consolidated sediments composed of clay, silt, sand
    (b in Fig. \ref{fig01}A) crop out. 
    Values for the factor of safety $F_{\mbox{s}}$ 
    are shown with a color ramp ranging from large values (green) to small values (red). 
    $F_{\mbox{s}}$ values smaller than unity (shown in blue) represent cells predicted 
    unstable by the model.
    }
  \label{fig02}
\end{figure}
% =================================================================== End Fig TWO%
and cover 523 km$^2$ (12.8\% of the catchment), 
for a total estimated landslide volume of 5.9$\times10^9$ m$^3$ \citep{Guzzetti:2008b}.

For the numerical experiments we selected the area in the UTRB where unconsolidated 
and poorly consolidated continental sediments crop out (green area in Fig. \ref{fig01}). 
This is the lithological complex where shallow landslides are more frequent in the study area
\citep{Cardinali:2000,Cardinali:2006,Guzzetti:2008b}, and where the TRIGRS conceptual
landslide scheme is best suited to model the stability/instability conditions of slopes forced 
by rainfall. The geotechnical properties for the soils in this lithological complex used for 
the numerical modeling are listed in line (b) of Table \ref{tab01}. 

% == MODEL DESCRIPTION ============================================================

\section{Distributed slope stability model}\label{sec:model}

As noted, we adopted TRIGRS version 2.0 \citep{Baum:2008}. The software implements a grid-based, 
spatially distributed slope stability model coupled with an infiltration model capable of simulating 
% ==================================================================== Fig THREE
\begin{figure}[!htp]
  \centerline{\hspace{0cm}
    \epsfxsize=0.45\textwidth\epsfbox{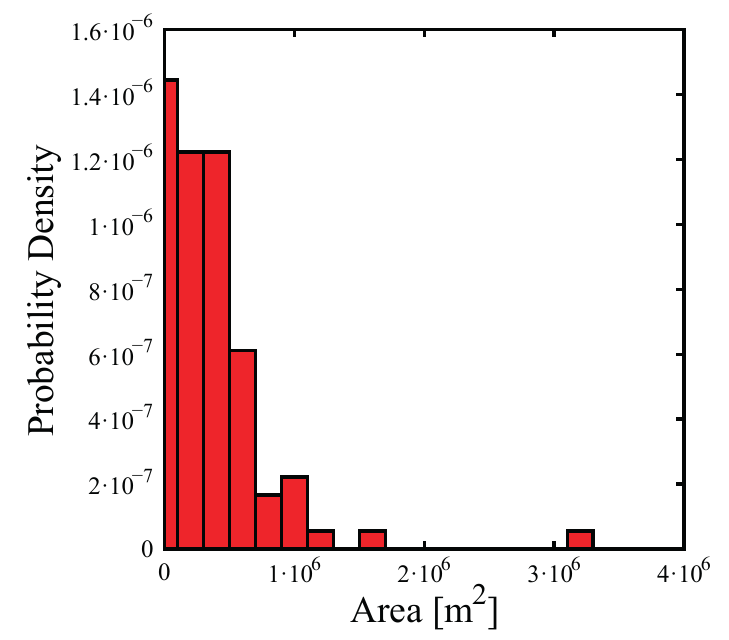}}
  \caption{
    Distribution of the area of the sub-basins resulting in well-defined
    rainfall thresholds, shown in Figs. \ref{fig04}, \ref{fig05} and \ref{fig06}.
    }
  \label{fig03}
\end{figure}
% =================================================================== End Fig THREE
the infiltration of rainfall in the terrain, modulating the stability/instability conditions 
of the individual grid cells \citep{Iverson:2000,Baum:2008,Godt:2008a,Godt:2008b}.

TRIGRS adopts a grid-based representation of a real landscape based on a DEM, and uses local 
terrain characteristics as an input for the solution of a system of equations whose output is the 
factor of safety $F_{\mbox{s}}$ \ie, a positive number representing the balance of the 
driving and the resisting forces acting in each grid cell.
For stable conditions, where the resisting forces exceed the driving forces, $F_{\mbox{s}} > 1.0$.
$F_{\mbox{s}} = 1.0$ represents the metastable condition where the driving and the resisting forces 
are equal, and $F_{\mbox{s}} < 1.0$ represents the physically unrealistic condition where the driving 
forces exceed the resisting forces, and the slope fails. 

For the modeling, TRIGRS exploits an infinite slope approximation \ie, a rigorous, lowest order 
approximation of a multi-dimensional landslide geometry. This approximation is adequate where $H<<L$, 
where $H$ is the depth of the slip surface and $L$ is the landslide linear dimension \citep{Iverson:2000}. 
An upper bound for $H$ can be identified on the basis of the lithological setting, where stronger rocks 
underly a weaker layer of soil or regolith. This is an assumption for shallow landslides for which 
depth is significantly smaller than width, or length. 
% ======================================================================= Fig FOUR
\begin{figure}[!htp]
  \vskip 0.1cm
  \centerline{\hspace{0cm}
    \epsfxsize=0.41\textwidth\epsfbox{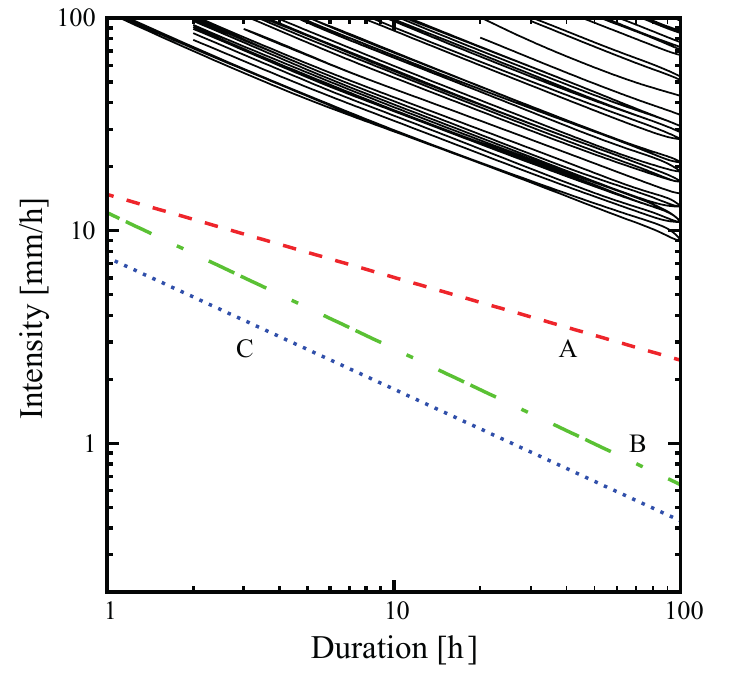}}
  \caption{    
    Rainfall intensity -- duration ($I$--$D$) thresholds.
    Colored lines are thresholds applicable to the UTRB obtained from: 
    A (red) $I$ = 14.82 $D^{-0.39}$ \citep{Caine:1980}, 
    B (green) $I$ = 12.17 $D^{-0.64}$ \citep{Brunetti:2010}, and
    C (blue) $I$ = 7.5 $D^{-0.62}$ \citep{Peruccacci:2012}. 
    Exponents for the black lines are in the range -0.54 $\pm$ 0.09,
    obtained fitting each threshold line with a power law, averaging 
    the exponents, and combining in quadrature the uncertainties given 
    by the fits.
  }
  \vskip -0.1cm
 \label{fig04}
\end{figure}
% ================================================================== End Fig FOUR
For simplicity, TRIGRS neglects all the lateral stresses and the inter-cell forces, and the stability 
of each grid cell is governed solely by the balance of the vertical component of gravity, against the 
resisting stress due to the basal Coulomb friction, plus the pore pressure \citep{Richards:1931}. 
Failure occurs at depth $Z$, measured vertically from the topographic surface, if at that depth 
\citep{Iverson:2000}: 
\beq
\label{effesse}
F_{\mbox{s}}\,=\,F_{\mbox{f}}\,+\,F_{\mbox{w}}\,+F_{\mbox{c}}\,=\,1\,.
\eeq
where, $F_{\mbox s}$ is the ratio of resisting to driving forces, and the individual components in 
Eq. (\ref{effesse}) can be written as a function of the local cell characteristics \citep{Iverson:2000}:
\beqy
\label{effeffe}
F_{\mbox{f}}&=&\tan \varphi / \tan \alpha\,,\\
\label{effedabliu}
F_{\mbox{w}}&=&\frac{-\psi(Z,t)\,\gamma_{\mbox{w}}\,\tan \varphi}{\gamma_{\mbox{s}}\,Z\,\sin \alpha \,\cos \alpha}\,,\\
\label{effeci}
F_{\mbox{c}}&=&\frac{c}{\gamma_{\mbox{s}}\,Z\,\sin \alpha \,\cos \alpha}\,,
\eeqy
where: $c$ is the soil cohesion, $\varphi$ is the soil internal friction angle, $\gamma_{\mbox{s}}$ is the 
wet soil unit weight, $\gamma_{\mbox{w}} = 9800$ N m$^{-3}$ is the unit weight of water, and $\alpha$ is the 
local angle of the terrain with respect to the horizontal. The model assumes that rainfall modulates the 
% ======================================================================= Fig FIVE
\begin{figure*}[!htp]
  \centerline{\hspace{0cm}
    \epsfxsize=1.05\textwidth\epsfbox{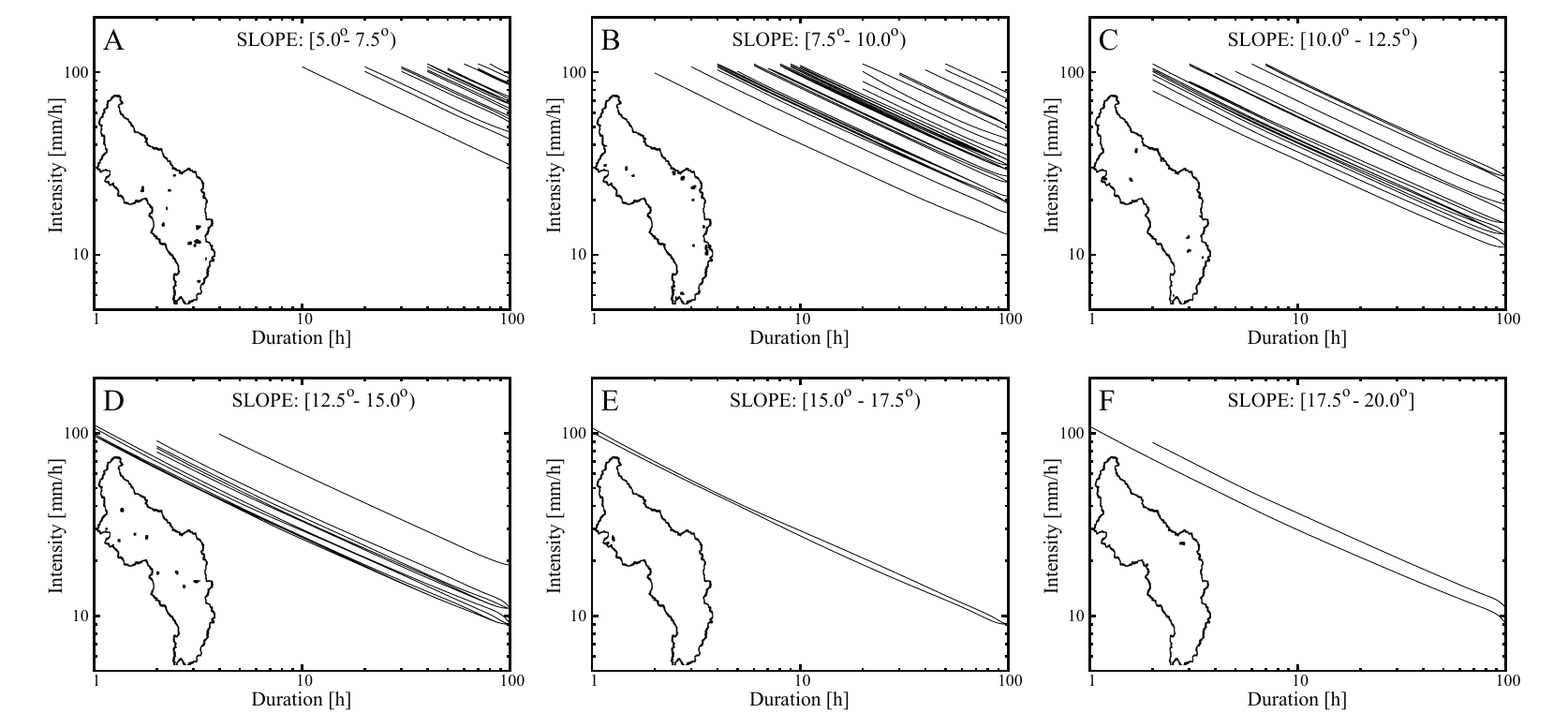}}
    \caption{
      Dependence of the modeled rainfall $I$--$D$ thresholds on the (mean) terrain 
      gradient for 91 sub-basins in the study area (b in Fig. \ref{fig01}A). 
      Slope increases from (A) to (F). Insets show geographical location of the 
      sub-basins for which the rainfall thresholds are shown in each plot. 
      Square bracket indicates value is included and round bracket indicates 
      value is not included.
    }
  \label{fig05}
\end{figure*}
% ==================================================================== End Fig FIVE
water table height and groundwater flow parallel to the slope, and that sub-surface flow is one-dimensional 
\citep{Iverson:2000}.
When rainfall hits the ground, due to water infiltration the driving force described by Eq. (\ref{effedabliu}) 
is modulated, $F_{\mbox{s}}$ varies with $Z$ and $t$, and $\psi(Z,T)$ is the pressure head as a function of depth 
$Z$ and time $t$. In Eq. (\ref{effedabliu}), the effects of variation is determined for each cell by solving 
the Richards equation \citep{Richards:1931}:
\beq
\label{richeq}
\frac{\partial \theta}{\partial t}\,=\,\frac{\partial}{\partial Z}
\,\left(K(\psi)\,\frac{1}{\cos^2 \alpha}\,\frac{\partial \psi}{\partial Z}\,-\,1\right)\,,
\eeq
where $\theta$ is the soil water content, and $K(\psi)$ is the soil hydraulic conductivity.

In  TRIGRS, Eq. (\ref{richeq}) is linearized and solved at discrete time steps and in the
vertical coordinate. The linearization procedure relies on the identification of two different 
time scales  \citep{Iverson:2000}. 
For each grid cell, $A$ is the upslope contributing area \ie, the cumulated area of all the 
upslope cells draining in the considered cell, and $D_0=K_{\mbox{s}}/S$ is the soil diffusivity 
in the cell, where $K_{\mbox{s}}$ is the saturated hydraulic conductivity (see Table \ref{tab01}) 
and $S$ is the specific water storage. The first time scale can be identified with $A/D_0$ as 
the time for lateral pore pressure transmission from the upstream area to the  grid cell. 
The second time scale is $H^2/D_0$, the time needed for pore pressure transmission from the 
surface to the depth $H$ \citep{Iverson:2000}. One can then build the length scale ratio 
$\varepsilon$:
\beq
\label{epsdef}
\varepsilon\,=\,\sqrt{\frac{H^2\,/\,D_0}{A\,/\,D_0}}\,=\,\frac{H}{\sqrt{A}}\,.
\eeq
Under the condition $\varepsilon << 1$, Eq. (\ref{richeq}) can be simplified identifying
long-term and short-term response terms \citep{Iverson:2000} used in the numerical implementation
of \citet{Baum:2008}. Eq. (\ref{epsdef}) is also used to identify the approximate limits of 
applicability of the model implemented by TRIGRS.
When rainfall intensity exceeds the local infiltration capacity, the excess water in the 
single grid cells is routed downslope to the nearest cells \citep{Baum:2008,Raia:2013}. 
TRIGRS accepts as input complex rainfall histories (\ie, spatially and temporally varying), 
permitting a realistic modeling of the slope stability/instability conditions driven by real 
rainfall events.

Fig. \ref{fig02} shows the result of the TRIGRS modeling for two representative sub-basins in the 
study area. The map portrays, superimposed on a shaded relief image showing the terrain morphology, 
the geographical distribution of the $F_{\mbox{s}}$ calculated using typical values for the mean 
% ======================================================================== Fig SIX
\begin{figure*}[!htp]
  \centerline{\hspace{0cm}
    \epsfxsize=1.05\textwidth\epsfbox{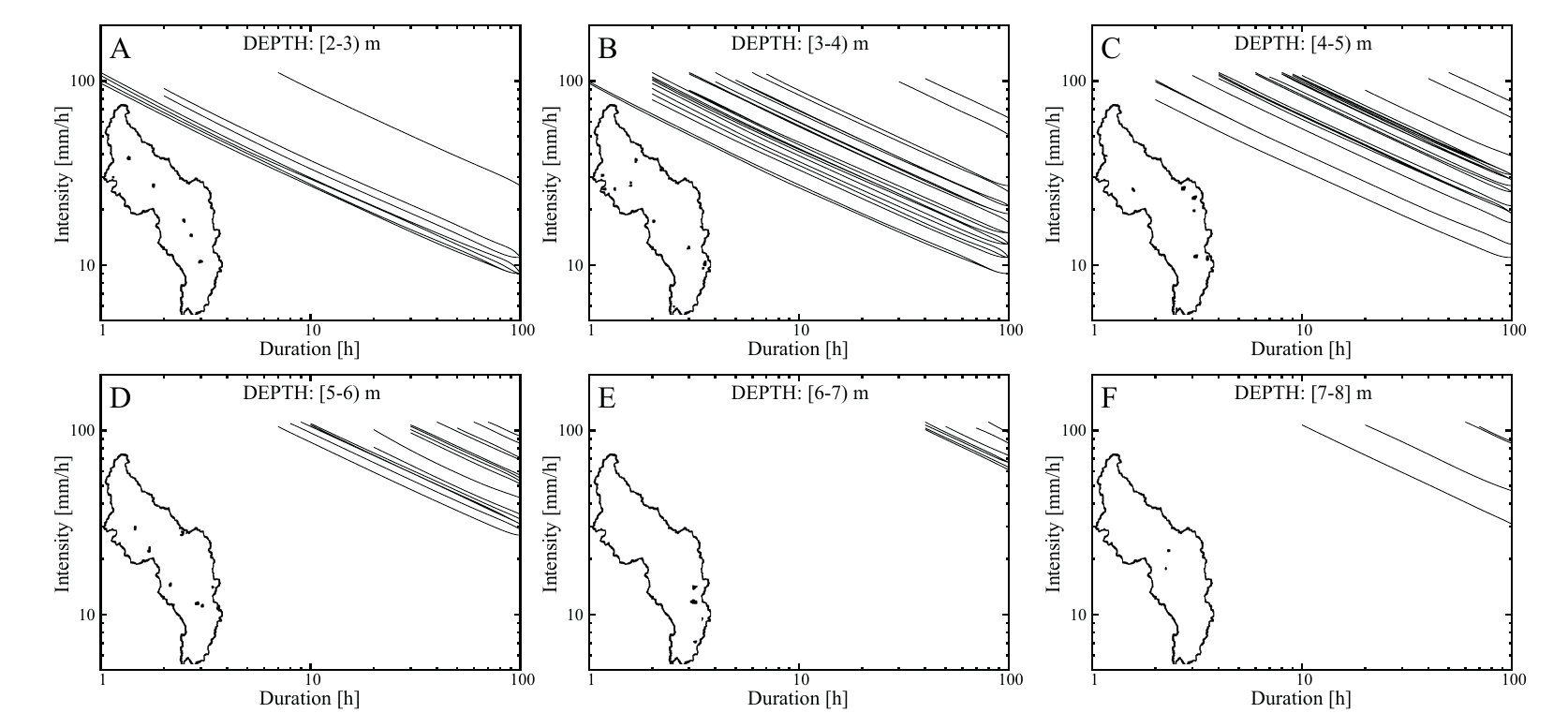}}
  \caption{
    Dependence on the modeled rainfall $I$--$D$ thresholds on the (mean) soil 
    depth for 91 sub-basins in the study area (b in Fig. \ref{fig01}A). 
    Soil depth increases from (A) to (F). Insets show geographical location 
    of the sub-basins for which the rainfall thresholds are shown in each plot. 
    Square bracket indicates value is included and round bracket indicates value 
    is not included.
    }
  \label{fig06}
\end{figure*}
% =================================================================== End Fig SIX
rainfall intensity and the rainfall duration. Shades of colours, from red ($F_{\mbox{s}} = 1$) to 
green, show the values of $F_{\mbox{s}}$ calculated for each grid cell. Single unstable cells, or 
clusters of unstable cells with $F_{\mbox{s}} < 1$ are shown in blue.

% == RAINFALL INTENSITY / DURATION =========================================== 
\section{Investigating the rainfall intensity -- duration dependence}\label{sec:iddep}

To investigate the relationship between the magnitude of the rainfall trigger (\ie, $I$ and $D$) 
we first partitioned the study area into sub-basins \ie, hydrological ensembles of slope units, 
where a slope unit is a hydrological region bounded by drainage and divide lines 
\citep{Carrara:1991,Guzzetti:1999}. To obtain the sub-basins we exploited the 25 $\times$ 25 m 
resolution DEM, and we used the \texttt{r.watershed} command in the GRASS GIS, release 7.0 
(www.grass.org). Sub-basins with a surface area smaller than 2,500 m$^2$ (\ie, four grid cells) 
were identified and excluded from the analysis. 
Fig. \ref{fig03} shows the statistical size-distribution of the slope units in the study area \ie,
where post-orogenic sediments crop out in the UTRB (green areas in Fig. \ref{fig01}).
Next, we forced the individual  sub-basins in the study area with a uniform rainfall of a given 
intensity $I$, for a given period of time $D$. 
We then studied the rainfall conditions that have resulted in unstable cells in the sub-basins. 
For simplicity, we considered only (nearly) mono-lithological sub-basins 
\ie, sub-basins for which at least 50\% of the area is covered by unconsolidated and poorly 
consolidated continental sediments. This reduced the complexity that the presence of multiple 
rock types in a sub-basin may introduce. We further considered unstable the sub-basins with at 
least 10\% of the grid-cells with $F_{\mbox{s}} < 1.0$. 
This is a reasonable assumption for rainfall-triggered landslides in the study area \citep{Cardinali:2006}.
The results depend only weakly on the proportion of cells in a sub-basin necessary to consider a sub-basin 
as unstable.

The first set of experiments was conducted adopting the following procedure. For each sub-basin, 
we started with a given (reasonable \eg, \citealt{Guzzetti:2007,Guzzetti:2008a}) set of $D$ and 
mean $I$ conditions, and we checked if these conditions resulted in landslides \ie, if at least 10\% 
of the grid-cells in the sub-basin had $F_{\mbox{s}} < 1.0$. 
Next, we increased $I$ maintaining $D$ constant, and we checked if the new rainfall conditions 
had resulted in landslides. The procedure was repeated for different, increasing rainfall durations. 
For the modeling, we used 
% =================================================================== Fig SEVEN
\begin{figure*}[!htp]
  \centerline{\hspace{0cm}
    \epsfxsize=0.9\textwidth\epsfbox{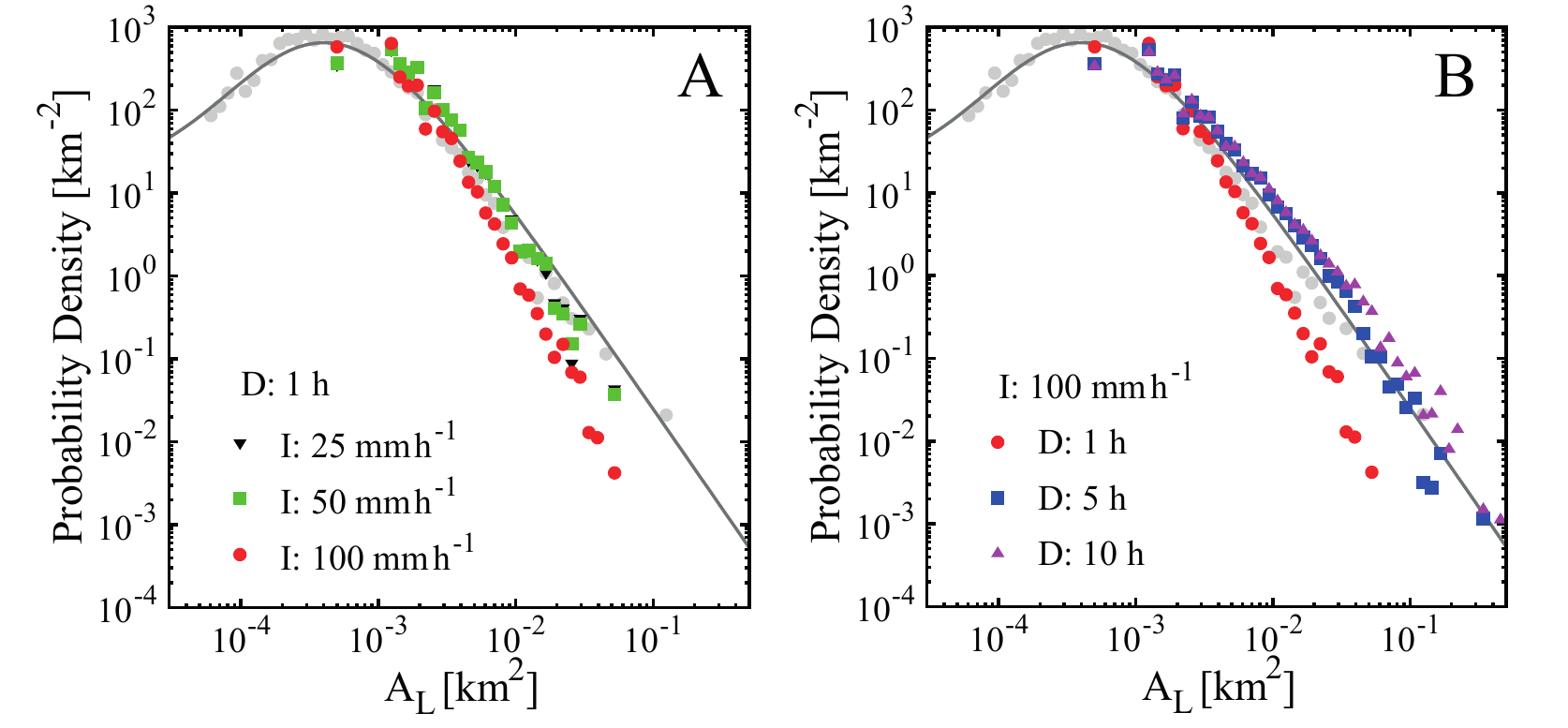}}
  \caption{
    Probability density of the area $A_{\mbox{\small L}}$ of the patches of connected 
    grid cells predicted as unstable by TRIGRS ($F_{\mbox{s}} < 1.0$).
    (A) shows dependence of the results on rainfall intensity.
    (B) shows dependence of the results on rainfall duration.
    See text for explanation.
    Grey dots show the probability density of the area of natural landslides 
    in the UTRB \citep{Cardinali:2000,Guzzetti:2008b}.
    Grey line shows the general probability density curve for event landslides 
    proposed by \citet{Malamud:2004}.
    }
  \label{fig07}
\end{figure*}
%% ============================================================== End Fig SEVEN
(i) \textit{I} = 1 mm h$^{-1}$ to \textit{I} = 200 mm h$^{-1}$ with steps of 2 mm h$^{-1}$, and  
(ii) \textit{D} = 1 hour to \textit{D} = 100 hours, with steps of 1 hour for the 1-10 h range, 
and steps of 10 hours for the 10-100 h range. At the end of the procedure, for each of the 
considered sub-basins we obtained a set of rainfall $(D,I)$ conditions that had resulted in 
(predicted) slope instabilities (10\% or more of the grid-cells with $F_{\mbox{s}} < 1.0$). We 
plotted these rainfall conditions in a $D,I$ plot, in log-log coordinates. The result is shown 
in Fig. \ref{fig04} where each black line represents the threshold above which a single sub-basin 
in the study area is unstable.

Visual inspection of Fig. \ref{fig04} reveals that the rainfall $(D,I)$ conditions that have 
resulted in unstable conditions in each sub-basin follow power laws that represent rainfall 
thresholds for slope instability in the sub-basins. 
The lines are equivalent to rainfall thresholds for possible 
landslide occurrence \citep{Guzzetti:2007}. Not all the sub-basins in the study area resulted 
in an instability threshold curve, because: 
(i) a number of sub-basins contained less than 50\% of continental, post-orogenic sediments, 
and were excluded from the analysis, 
(ii) the percentage of unstable cells in a sub-basins was less than 10\%, regardless of the rainfall conditions, 
and where not considered as failed slopes, and
(iii) the instability threshold curve was outside of the ranges of rainfall intensity and duration 
considered for the modeling. 

Further inspection of Fig. \ref{fig04} reveals that the threshold lines are well defined, and 
obey distinct power law trends for a  significant range of rainfall durations 
($1 \leq D \leq 100$ hours) known to initiate landslides \citep{Guzzetti:2007,Guzzetti:2008a}. 
Significantly, the slope of the threshold curves is in agreement with the slope of empirical 
rainfall thresholds for possible landslide occurrence established for central Italy 
\citep{Peruccacci:2012} and for Italy \citep{Brunetti:2010}, considering the uncertaintly 
associated with the empirical thresholds \citep{Peruccacci:2012}.
The position of the threshold curves in the $D,I$ plane depends on the location and geomorphological 
characteristics of the individual sub-basins,
but all of the threshold curves lay above the empirical thresholds proposed in the literature 
and applicable to the study area (\eg, \citealt{Caine:1980,Brunetti:2010,Peruccacci:2012}). 
This is because empirical thresholds are defined as lower 
boundaries of $I$, $D$ conditions in a large area, while in our case each curve corresponds 
to a specific (local) morpho-lithological setting.
We stress that the result was obtained using reasonable 
values for the model parameters, but without any attempt to fine-tune the parameters, and specifically 
the geotechnical parameters (Table \ref{tab01}). Also, no attempt was made to optimize the modeling 
% ==================================================================== Fig EIGHT
\begin{figure*}[!htp]
  \centerline{\hspace{0cm}
    \epsfxsize=0.9\textwidth\epsfbox{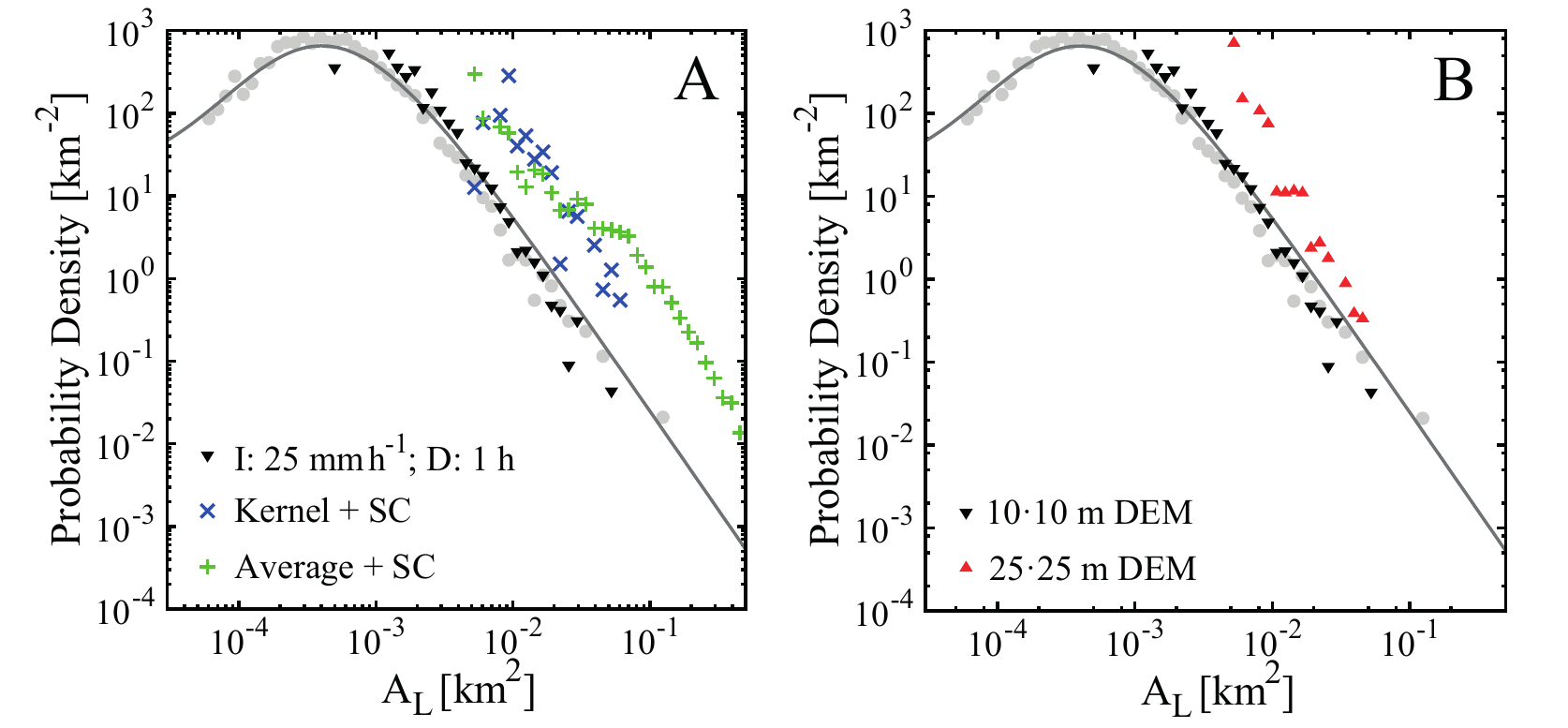}}
  \caption{
    Probability density of the area $A_{\mbox{\small L}}$ of the patches of connected
    grid cells predicted as unstable by TRIGRS ($F_{\mbox{s}} < 1.0$).
    (A) shows dependence on clustering strategy. 
    (B) shows dependence on DEM resolution.
    Grey dots show the probability density of the area of natural landslides 
    in the UTRB \citep{Cardinali:2000,Guzzetti:2008b}.
    Grey line shows the general probability density curve for event landslides 
    proposed by \citet{Malamud:2004}.
  }
  \label{fig08}
\end{figure*}
%% =================================================================== End Fig EIGHT
parameters using \eg, a probabilistic modeling approach \citep{Raia:2013}. 
We maintain that this indicates that the scaling behavior of the rainfall conditions responsible
for shallow landslides in the UTRB emerged from the physical modeling, and is  dependent on the 
local geomorphological setting and the rainfall trigger.

To further investigate the position of the obtained threshold curves in the $D,I$ plane with the local 
terrain slope and the soil depth, two key variables that control the modeling \citep{Iverson:2000,Baum:2008}, 
we separated the threshold curves based on classes of terrain slope (Fig. \ref{fig05}), and soil depth 
(Fig. \ref{fig06}) in the sub-basins. Visual inspection of Fig. \ref{fig05} reveals that the $I\mhyphen D$ 
threshold curves for sub-basins characterized by steep slopes are lower that the thresholds for sub-basins 
characterized by gentle slopes. This was expected \citep{Iverson:2000,Baum:2008}. Analysis of Fig. \ref{fig05} 
indicates that thinner soils fail with less severe rainfall conditions (lower thresholds) than the thicker soils. 
This was also expected \citep{Iverson:2000,Baum:2008}. We attribute the dispersion of the threshold lines in the 
different classes of terrain slope and soil depth to natural variability \ie, to the terrain variability that 
characterizes the sub-basins in the study area. In Figs. \ref{fig05} and \ref{fig06} the maps show the 
geographical location of the sub-basins for which the threshold lines shown in the same panel were calculated. 
Inspection of the maps reveals that the sub-basins are distributed throughout the study area, with no 
geographical trend or bias. This is an indication that the results are independent of a specific location 
or geomorphological setting.

% == LANDSLIDE SIZE ========================================================
\section{Investigating the distribution of the size of the landslides}\label{sec:sizes}

We determined the probability density of the area of the patches of terrain predicted 
as unstable by TRIGRS, and we compared the probability density of the patches to the probability density 
of natural landslides in the UTRB \citep{Malamud:2004,Guzzetti:2008b}. 
We defined a patch of unstable terrain a cluster of contiguous grid cells that individually 
have  $F_{\mbox{s}} < 1.0$. To identify the unstable grid cells, we ran TRIGRS with a fixed rainfall duration 
$D$ = 1 hour, increasing the rainfall intensity $I$ from 25 to 100 mm h$^{-1}$. Next, we repeated the 
calculations with a fixed rainfall intensity $I$ = 100 mm h$^{-1}$, and we varied $D$ 
from 1 to 10 hours. The remaining model parameters were kept constant for all the model runs (see 
Table \ref{tab01}). Results are summarized in Fig. \ref{fig07}.
For the experiment shown in Fig. \ref{fig07} we used the finer resolution (10 
% ======================================================================== Fig NINE
\begin{figure*}[!htp]
  \centerline{\hspace{0cm}
    \epsfysize=0.34\textwidth\epsfbox{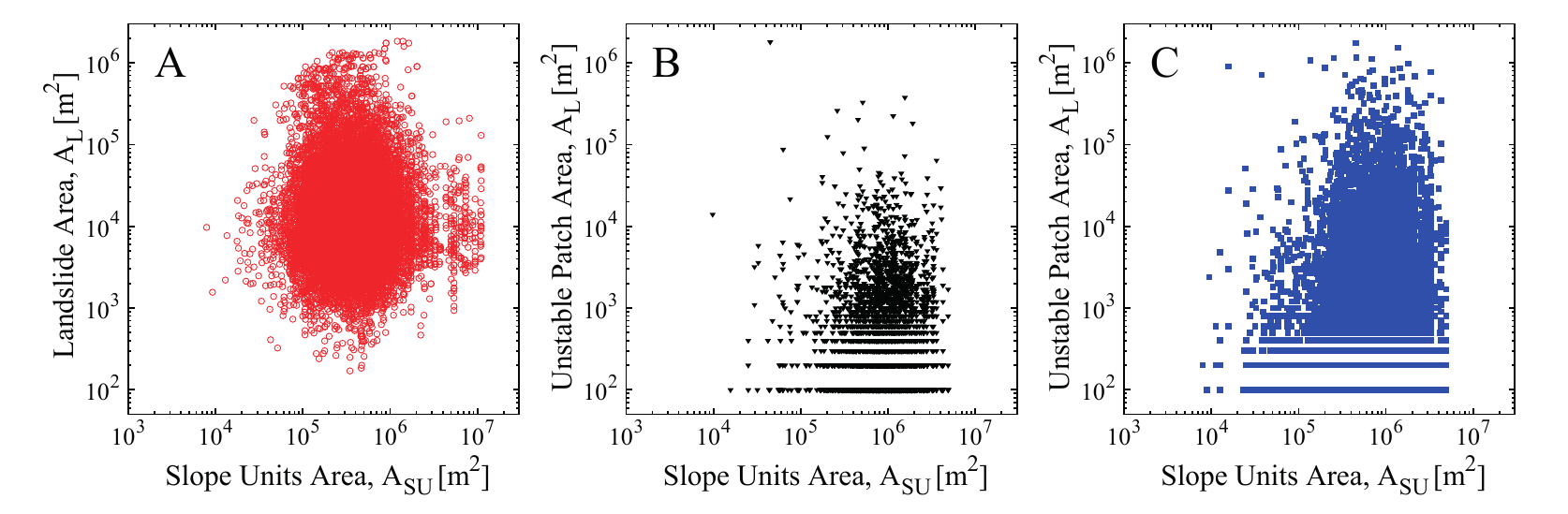}}
  \caption{
    Relationship between the area of the slope units $A_{\mbox{\tiny SU}}$,
    and the area of the individual landslides $A_{\mbox{\tiny L}}$ in the UTRB,
    or the area of the patches of connected grid cells predicted as unstable by TRIGRS,
    where post-orogenic sediments crop out (b in Fig. \ref{fig01}A).
    (A) shows results for landslides mapped by \citet{Guzzetti:2008b} in the UTRB;
    (B) shows results of TRIGRS using $D$ = 1 h, $I$ = 25 mm h$^{-1}$, and 
    (C) hsows results of TRIGRS using $D$ = 5 h, $I$ = 100 mm h$^{-1}$.
    The probability distribution for the results in (B) and (C) are shown in Fig. 
    \ref{fig07}; some DEM resolution effect is clearly seen in the lower region of 
    both plots. 
  }
  \label{fig09}
\end{figure*}
%% =================================================================== End Fig NINE
$\times$ 10 m) DEM \citep{Tarquini:2007,Tarquini:2012}. 

Visual inspection of Fig. \ref{fig07} reveals a good agreement between the probability density 
of the area of the patches of unstable terrain, and the corresponding probability density 
of event landslides in the UTRB \citep{Cardinali:2000,Cardinali:2006}, and with the general 
probability density curve for event landslides prosed by \citet{Malamud:2004}. 
The agreement is best for medium to low rainfall intensities ($I$ = 25 to 50 mm h$^{-1}$) 
and for medium rainfall durations ($D$ = 5 h).
These rainfall conditions are known to initiate landslides in the UTRB 
and in similar physiographic regions in Umbria \citep{Cardinali:2006} and in Central Italy \citep{Peruccacci:2012}.
Higher rainfall intensities and shorter rainfall periods result in a poorer 
match between the modeled and the empirical probability densities. 

For small and very small unstable patches which we identify as single landslides, with area
$A_{\mbox{\small L}} < 10^{-3}$ km$^2$, the probability density is lower than prescribed by a 
simple negative power law. Interestingly, the reduced density occurs at the same size 
$A_{\mbox{\small L}}$ for which the probability density of the natural landslides exhibits 
a distinct rollover, which was recognized to be a physical characteristic of populations 
of event landslides \citep{Malamud:2004}, and was attributed to the mechanical properties 
of the soil and regolith where the small landslides occur \citep{Katz:2006,Stark:2009}. 
However, the smooth rollover for small and very small landslides typical of event landslides 
is not clearly reproduced by the model. We cannot exclude a priori that the result is related 
to the strategy used to decide the size of the patches of instability, or to the resolution 
of the DEM used for the modeling. 

To investigate this possibility, we tested different strategies to identify the patches of 
unstable terrain, and we studied the effects of the different strategies on the probability 
density of the area of the patches. The simplest strategy consisted in considering all the 
grid cells in a sub-basin with $F_{\mbox{s}}$ $<$ 1.0 that shared a boundary or a corner as 
belonging to a single landslide.
In this strategy, a single unstable cell not connected to any other unstable grid cell represents 
an unstable patch with the smallest possible area, $A_{\mbox{\small L}}$  = 100 m$^2$ for a 10 $\times$ 
10 m DEM. An alternative strategy consisted in smoothing the grid map of the computed $F_{\mbox{s}}$ 
values prior to searching for the contiguous cells. For the purpose, we passed a 3 $\times$ 3 kernel 
over the grid map, and decided that the central pixel was unstable ($F_{\mbox{s}}$ $<$ 1.0) if at least 
two (of the nine) cells in the kernel had $F_{\mbox{s}}$ $<$ 1.0. Another strategy consisted in passing 
the same 3 $\times$ 3 kernel over the $F_{\mbox{s}}$ grid map, and in deciding that the central pixel 
is unstable if it has $F_{\mbox{s}}$ $<$ 1 and if the average $F_{\mbox{s}}$ value of the nine pixels 
in the kernel was 
%$<$ 
smaller than 5, a figure determined heuristically through an iterative procedure.
The first strategy provided the best result (Fig. \ref{fig07}A), with the other strategies 
deviating significantly from the landslide field data. For the first strategy, the agreement 
between the area of the modeled patches and the area of the event landslides in the UTRB is 
best for $2 \times 10^{-3} < A_{\mbox{\small L}} \leq  2 \times 10^{-2}$ km$^2$.
The density deviate slightly for larger areas ($A_{\mbox{\small L}} > 2 \times 10^{-2}$ km$^2$).
We attribute the deviation to the fact that very large unstable patches correspond to large, 
deep-seated landslides that are not modeled by TRIGRS. 

We investigated the dependence of the model results to the ground resolution of the DEM. For 
the purpose, we repeated the model runs using the coarser 25 $\times$ 25 m DEM. The results 
are shown in Fig. \ref{fig08}B, and were obtained using the same rainfall intensity, 
rainfall duration, and geotechnical parameters, changing the resolution of the DEM. Inspection 
of the plot shows that the probability density of the area of the patches of unstable grid 
cells obtained using the finer resolution (10 $\times$ 10 m) DEM follows nicely the probability 
density of the area of event landslides in the UTRB, for $A_{\mbox{\small L}} \leq 2 \times 10^{-2}$ 
km$^2$. Again, we attribute the observed deviation for larger areas to the fact that the very 
large patches correspond to deep-seated landslides that are not modeled by TRIGRS. Instead, the 
results obtained using the coarser resolution (25 $\times$ 25 m) DEM follows a steeper power law 
trend that deviates from the probability density of the natural landslides in the UTRB 
\citep{Malamud:2004}. 
We repeated the simulations with the coarser DEM using larger intensity and duration values,
obtaining curves with a lower slope, with a limit to the correct one. However the resulting 
probability densities as a function of $A_{\mbox{\small L}}$ (not shown in Fig. \ref{fig08}) 
were higher than the curves obtained using the finer resolution DEM. We conclude that the 
resolution of the coarser DEM is insufficient to model the stability/instability conditions 
of shallow landslides in the study area. 

Lastly, we investigated the relationship between the size of the patches predicted unstable 
by the TRIGRS model (Fig. \ref{fig03}) and the size of the slope units  in the UTRB (Fig. 
\ref{fig09}). We find that the size distribution of the predicted unstable patches covers 
the same size range of the landslides mapped in the UTRB, and that the relation between 
the size of the unstable patches and the size of the slope units depends on the (synthetic) 
$I\mhyphen D$ rainfall conditions. The differences between the modelled and the empirical 
distributions have multiple causes, including the method used to cluster the unstable cells, 
and the fact that our modeling results were obtained for the portion of the UTRB where the 
post-orogenic sediments crop out (area b in Fig. \ref{fig01}).

\section{Comparison with other modeling approaches}\label{sec:soc}

We attempted a comparison between the results obtained in our study area using the physically-based 
model TRIGRS \citep{Baum:2008}, and other attempts conducted to exploit numerical, simplified (\ie, 
``toy"), or conceptual models to simulate the rainfall conditions responsible for slope failures, 
and the frequency-size distribution of the landslides. We concentrate on simplified, conceptual 
models, and we exclude the use of physically-based geo-mechanical models such as those proposed
by \citet{Katz:2006} or by \citet{Stark:2009} to describe the frequency-size statistics of slope 
failures, or other physically-based spatially distributed slope stability models \eg, SHALSTAB  
\citep{Montgomery:1994,Dietrich:1998}, SINMAP \citep{Pack:1988}, SHETRAN \citep{Burton:1998}, 
STARWARS \citep{Malet:2005}, GEOtop-FS \citep{Rigon:2006,Simoni:2008}, and CHLT \citep{vonRuette:2013}.

We are not aware of any attempt to describe the scaling behaviour of rainfall thresholds for possible 
landslide occurrence executed using simplified or conceptual models. The results obtained in this work 
represent a preliminary attempt to provide a physical background to the empirical rainfall thresholds 
for the prediction of possible landslide occurrence. This is important for the application of rainfall 
thresholds in landslide warning systems \citep{Chleborad:2003,Aleotti:2004,Godt:2006,Guzzetti:2008a,
Brunetti:2009b}. 

Investigators have attempted to simulate the self-similar scaling behaviour of landslide size using 
Cellular Automata (CA) models (see \eg, \citealt{Hergarten:1998,Hergarten:2000b,Hergarten:2000,
Hergarten:2002,Piegari:2006a,Piegari:2006b,Piegari:2009}). 
The interest in the use of CA models to investigate landslide sizes is motivated largely by the analogy
between landslides in a landscape and avalanches in the  ``sandpile" model, the first theoretical and 
practical realization of a Self-Organized Critical (SOC) system \citep{Bak:1987,Bak:1988}. In the sandpile 
model, the meta-stable condition is controlled by a critical slope, which gives rise to relaxation of the 
system \ie, movements of single or multiple grains between cells that result in an avalanche-like behaviour. 
SOC systems are implemented numerically in a number of variants, choosing different dynamical variables or 
combination of variables to simulate the interactions between the slope and its time-dependent weakening 
\citep{Hergarten:2000}. Typically, the critical conditions are controlled by conservative or dissipative 
relaxation rules and by the boundary conditions, leading to an SOC behaviour. 
When applied to the investigation of the statistics of landslide sizes, a limitation of CA models consists 
in the fact that the exponents controlling the scaling of the distributions obtained through the CA models 
are significantly smaller (\eg, \citealt{Hergarten:2000,Hergarten:2002,Hergarten:1998,Hergarten:2000b})
than the corresponding exponents obtained modeling empirical data taken from landslide inventory maps
(\eg, \citealt{Pelletier:1997,Stark:2001,Malamud:2004,VanDenEeckhaut:2007}). 
The frequency distributions obtained from CA models are also very sensitive to the rules used to 
set up and run the models, and to the initial and boundary conditions \citep{Piegari:2006a,Piegari:2006b,
Piegari:2009}. \citet{Guzzetti:2002} hypothesized that an inverse cascade model could explain the power-law 
behavior of natural landslides and of SOC models of slope instability, but did not demonstrate a link between 
the scaling of the empirical and modeled distributions. 
In a recent paper, \citet{Hergarten:2012} proposed a criticality-inspired numerical model to explain the
frequency-size distribution of large rockfalls, and to estimate the size and frequency of the largest possible 
failures in a region. The model uses a DEM to calculate local terrain slope, and exploits random perturbations 
to mimic natural fractures that control the location and size of the rockfalls. Results obtained for the Alps 
revealed that the frequency distribution of the modelled failure volumes obeys a power-law with a scaling 
exponent in the range of the known empirical values \citep{Brunetti:2009a}, and exhibits a distinct ``rollover'' 
for small size failures.
 
\citet{Chen:2011} proposed an alternative way for describing the scaling properties of landslide sizes 
exploiting non-extensive statistical mechanics, or Tsallis statistics \citep{Tsallis:1988,Tsallis:1999,
Tsallis:2011}, a formal extension of the Boltzmann-Gibbs statistics. This approach relies on the postulate 
that a system composed by two independent statistical systems has an entropy given by the entropies of the 
two subsystems plus a correction term dependent on a parameter that controls the degree of non-additivity. 
Despite criticisms \citep{Lee:2012,Chen:2012}, the approach proved capable of describing quantitatively the 
frequency-size distributions typical of empirical landslide datasets, including the scaling exponents and 
the ``rollover'' observed for small landslide sizes \citep{Malamud:2004}. When applied to simulate the 
frequency-size statistics of landslides, the postulate of non-additivity of entropy implied by the Tsallis 
statistics is interpreted with the redistribution of the soil materials in different (larger) volumes when 
a landslide occurs, which increases the entropy in the final system composed of a different distribution 
of the initial sub-systems. The advantage of the approach is that the statistical distribution of the 
landslide size can be derived in an analytic way, starting from general principles \citep{Chen:2011}. 

As discussed before, our results indicate that a relatively simple, physically-based model for 
simulating the stability/instability conditions of natural slopes forced by rainfall, when applied 
to a sufficiently large area, is for reproducing well the frequency--area distribution of the 
natural landslides in the same area. This proves that the observed self-similar scaling behaviour of 
the landslide areas, and its deviation from the main trend observed for small landslides, are the 
results of physical characteristics of the landscape and of the slope instabilities in it. 

Further, application of the physically-based model revealed the presence of both scaling properties
of landslides \ie, of the intensity of the driving forces (the mean rainfall intensity and the rainfall 
duration responsible for the slope instabilities) and of the magnitude of the consequences (the relative 
proportion of landslides of different sizes). The result was obtained selecting reasonable values for the 
model parameters, without changing or optimizing them. 
This result indicates a functional link between rainfall and landslides, and opens to the possibility to 
quantify the link. 

TRIGRS reproduced well the statistical distribution of landslide areas and the rainfall conditions that 
have resulted in landslides in our study area. This is not a trivial result, because the landslide scaling 
properties captured by the numerical model are not keyed in the equations governing the model. We maintain 
that the fact that the distribution of landslide size and the rainfall thresholds for landslide occurrence 
are reproduced well by the model is evidence that the model uses the correct dynamics and variables. Our 
results open to the possibility of establishing rainfall thresholds for 
%% =================================================================== TABLE ONE
\begin{table*}[!htp]
  \caption{
    Geotechnical properties for the soils in the lithological complexes cropping out in the 
    UTRB (Fig. \ref{fig01}; \citealp{Cardinali:2001}).
    $c$ is the soil cohesion, 
    $\varphi$ is the soil internal friction angle, 
    $\gamma_{\mbox{s}}$ is the wet soil unit weight, 
    $D_0$ is the soil diffusivity, and 
    $K_{\mbox{s}}$ is the soil saturated hydraulic conductivity. 
    See Section \ref{sec:model} for details.
    The codes (a) to (e) match the corresponding color codes in Fig. \ref{fig01} for the different 
    lithological complexes.}
    \vskip3mm
    \centering
    \begin{tabular}{ccccccc}
      \hline
      Unit &  $\,\,\,\,c$   & $\,\,\,\,\varphi$  & $\,\,\,\,\gamma_{\mbox{s}}$    &$\,\,\,\,D_0$              & $\,\,\,\,K_{\mbox{s}}
      $ \\   & [kPa]         &   [deg]               & [N m$^{-3}$]             &  [m$^2$ s$^{-1}$]   &    [m s$^{-1}$]\\      
      \hline          
      a &    2.5   &   10.0    & 15,000    & 4.7$\cdot10^{-3}$   & 1.0$\cdot 10^{-4}$ \\  
      \textbf{b} &   \textbf{3.0}   &   \textbf{15.0}    & \textbf{15,000}    & \textbf{4.7$\mathbf{\cdot10^{-3}}$}  
      & \textbf{1.0$\mathbf{\cdot 10^{-4}}$}\\ 
      c &  50.0   &   25.0    & 15,000    & 8.3$\cdot10^{-6}$   & 1.0$\cdot 10^{-6}$ \\  
      d &  50.0   &   30.0    & 15,000    & 8.3$\cdot10^{-6}$   & 1.0$\cdot 10^{-6}$ \\  
      e &  75.0   &   35.0    & 22,000    & 8.3$\cdot10^{-6}$   & 1.0$\cdot 10^{-6}$ \\ 
      \hline
    \end{tabular}
    \label{tab01}
\end{table*}
% ======================================================================== END TABLE
landslide occurrence for individual 
slopes or groups of slopes, reducing the geographical uncertainty associated with empirical rainfall thresholds, 
which are most commonly defined for large to very large areas \citep{Guzzetti:2007,Guzzetti:2008a,Peruccacci:2012}.
A reduced geographical uncertainty of the thresholds may improve the performance of landslide warning systems 
based on rainfall thresholds. Further, the results open to the possibility of predicting the magnitude of 
landslide triggering meteorological events, and the impact of rainfall induced landslides on the landscape.
Accurate information on the expected impact of landslides may be important for landscape evolution studies,
and for improved landslide hazard and risk assessments.  

% == CONCLUSIONS =====================================================
\section {Conclusions} \label {sec:concl}

We used the consolidated Transient Rainfall Infiltration and Grid-Based Regional Slope-Stability 
analysis code (TRIGRS, version 2.0; \citealp{Baum:2008}) to investigate two scaling properties 
of landslides, namely: 
(i) the rainfall intensity -- duration ($I\mhyphen D$) conditions responsible for slope instability 
and landslide occurrence \citep{Guzzetti:2007,Guzzetti:2008a}, and 
(ii) the probability density of the area of rainfall induced landslides \citep{Stark:2001,Malamud:2004}. 
Our results, obtained for a significantly large area in Central Italy where landslides are abundant 
and frequent \citep{Cardinali:2001,Cardinali:2006,Guzzetti:2008b}, 
indicate that the physically based model was capable of reproducing the two landslide scaling behaviors. 
More specifically, the numerical modeling revealed the followings:

\begin{itemize}

\item The scaling exponent of the power-law curves describing the mean rainfall intensity vs. rainfall 
duration conditions that produced slope instability in the studied (quasi) mono-lithological sub-basins,
where unconsolidated and poorly consolidated sediments crop out was almost identical to the scaling 
exponent of empirical $I\mhyphen D$ rainfall thresholds for possible landslide occurrence defined for 
Central Italy \citep{Peruccacci:2012}, and for Italy \citep{Brunetti:2010}. The finding is important 
because it reconciles the physically-based and the statistically-based approaches to the prediction of 
rainfall induced landslides \citep{Crosta:2003,Aleotti:2004,Guzzetti:2007,Guzzetti:2008a}. This is 
relevant for landslide warning systems based on rainfall thresholds \citep{Chleborad:2003,Aleotti:2004,
Godt:2006,Guzzetti:2008a,Brunetti:2009b,BachKirschbaum:2012}.

\item The probability density distribution of the area of the patches of grid cells predicted as 
unstable by the TRIGRS model (\ie, grid cells with $F_{\mbox{s}} < 1.0$) in the study area 
was similar to the frequency density of the area of natural event landslides in the UTRB 
\citep{Cardinali:2001,Guzzetti:2008b}, and 
to the general probability density distribution for event landslides proposed by \citet{Malamud:2004}, 
for areas $A_{\mbox{\small L}} >  2 \times 10^{-2}$ km$^2$. For smaller areas, the probability density of 
the patches of unstable cells deviates from the power law trend. This is similar to the ``rollover" 
observed in statistically complete populations of event landslides \citep{Malamud:2004}. 
The finding is relevant because it provides a physical basis to the statistics of landslide areas 
in a large and geomorphologically complex region.
\end{itemize}

We consider the results of this work as a starting point for investigating the underlying mechanisms 
giving rise to the scaling properties of landslide phenomena.

% == ACKNOWLEDGEMENTS ==================================================
\section*{Acknowledgments} 

We thank R.L. Baum and J.W. Godt (USGS) for valuable discussion.
We are grateful to the Editor (T. Oguchi), S. Hergarten and an anonymous reviewer for 
their constructive comments. 
MA was supported by grants provided by the Regione Umbria, under contract \textit{POR-FESR
Umbria 2007-2013, asse ii, attivit\`a a1, azione 5}, 
and by the \textit{Dipartimento della Protezione Civile}, Italy.

% == REFERENCES =========================================================
%%\section{References} 
\bibliographystyle{elsarticle-harv}

%------------------------------------------------------------------------------%
%------------------------------------------------------------------------------%
\end{document}